# The artist libraries project

**Félicie Faizand de Maupeou[1]\***

1 Université Paris Nanterre, France

*Corresponding author: feliciedemaupeou@gmail.com

**Abstract**

The creation of the *Artist Libraries Project* was sparked by the observation that artist libraries are still not well known, yet many art historians are interested in this archive for the value it adds to understanding the person behind the artist and his or her creative process. The problem is that these libraries are rarely physically preserved. To remedy this dispersion, we built an online database and a website www.lesbibliothequesdartistes.org that house this valuable source in the form of lists of books and their electronic versions. First data on Monet's library was made available, and several additional artist libraries from the 19th and 20th centuries are on the way for 2019. By gathering all these bibliographical data in a central database, it's possible to explore one library and to compare several. This article explains how we built the database and the website and how the implementation of those IT tools has raised questions about the use of this resource as an archive on the one hand, as well as its value for art history on the other.

## I CONTEXT

Following the famous sentence of Cezanne, *Monet ce n'est qu'un œil mais bon dieu quel œil!* the historiography of impressionism built the idea that the members of this movement were almost completely disconnected from their context and their time. Even if the thinking on this has evolved lately, many studies still ignore Monet's inspirations beyond just nature, and instead focus narrowly on his concern with painting, color and pattern, as if he was divorced from time and, especially, from any theoretical thought. Monet himself certainly promoted this image, claiming to abhor theory, and disputing that everyone has his own calling; his was to paint, not to comment on his painting. It is indeed true that Monet never wrote a theoretical text and that he was always reluctant to analyze his own works and his technique. As early as 1895, he categorically declared: "[...] j'ai horreur d'être mis en scène… c'est déjà bien assez de livrer au public ce que l'on fait sans l'assommer de ce que l'on pense[1]". In the same vein, he answers to Louis Vauxcelles, in 1900, about a questionnaire on the technique of painters to be published in *Le Figaro*: « Je comprends très bien l'intérêt que cela peut avoir pour vous, je le reconnais et serai le premier à lire des réponses qui vous seront adressées. Mais pour moi, je m'en tiens à mes pinceaux[2]". He maintains his point of view until the end of his life; a few months before he dies, in March 1926, he wrote: "[...] j'ai toujours eu horreur des théories, enfin que je n'ai que le mérite d'avoir peint directement devant la nature en cherchant à rendre mes impressions devant les effets les plus fugitifs, et je reste désolé d'avoir été la cause du nom donné à un groupe dont la plupart n'avait rien d'impressionniste…[3]".

---

[1] "I hate being put on stage ... it is already plenty enough to give to the public what one makes without smothering them with what one thinks", Letter W 1308 to an unknown address, Giverny, June 26, 1895. W refers to Daniel Wildenstein's *Monet: catalogue raisonné*, [1974]., Köln-Lausanne, Taschen -Wildenstein Institute, 1996,5 vol.

Nevertheless, it is difficult to embrace the argument that Monet was divorced from time and thought when we consider the intellectual, social, political and scientific context in which he lived in the late 19[th] century and turn of the 20[th] century. It is even less credible when we consider how revolutionary the impressionist movement was in the history of landscape painting. The presence of literary men and politicians in his entourage suggests that Monet was in fact strongly anchored in his time. More important, we can peer into his library as a source to understand the man behind the painter. There is origin of the *Artist Libraries Project*.

Monet's library has been conserved in the second studio of the painter. This building now houses the administrative offices of the Monet Foundation and is therefore not included in the public tour. An initial study on his library led to a book [Le Men et all, 2013]. This book is an anthology of texts taken from the books of the library, together with illustrations, either taken from these books, or from Monet's or his friends' work. This book also includes a bibliography of titles in the artist's library, some of which contain author dedication pages. This aroused interest among researchers and compelled us to expand our work to all artists' libraries. The problem is that those libraries are rarely physically preserved. Therefore, they require a great deal of investigation, which means reconstitution and cross-referencing of several sources, such as correspondence, workbooks or testimony of his close relatives. Such research is mostly conducted individually by researchers who adopt a monographic approach: a researcher interested in an artist finds his library and decides to study it as a new source to learn more about his personality or his work. Those lists of books are thus disseminated in monographic studies, published or unpublished academic works, articles, inventories of libraries... One of the major challenges of this project is to remedy this dispersion. To achieve this, digital technology, i.e. the creation of an online database, has appeared the best option, for several reasons. It makes it possible to handle large amounts of data, to cross-search by artist, work, author..., to enrich data thanks to interoperability and to browse digital versions. The launch of the Labex *Les Passés dans le present[4]*, hosted by Paris Nanterre University, has offered a proper field of production. The website www.lesbibliothequesdartistes.org was launched in Jan 2018 to house lists of works and electronic versions of works. First data on Monet's library was made available, and several additional artist libraries from the 19[th] and 20[th] centuries are on the way for 2019. By gathering all these bibliographical data in a central database, it's possible to explore one library and to compare several.

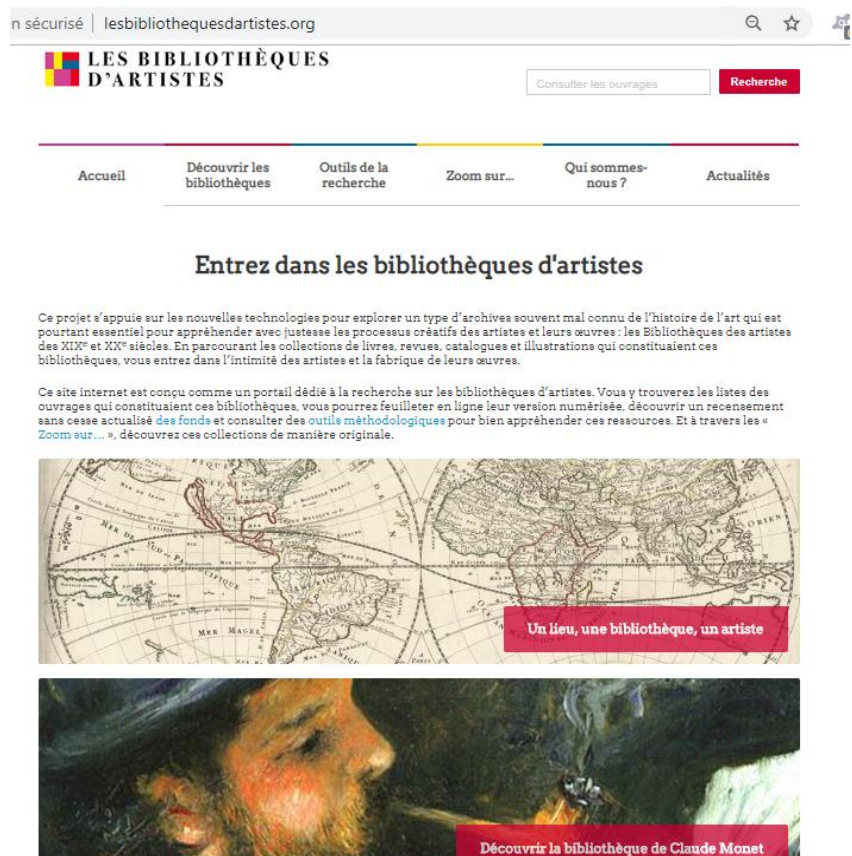

Figure 1. Home page of www.lesbibliothequesdartistes.org

## II THE ARTIST'S LIBRARY: AN IMPORTANT AND YET UNKNOWN RESOURCE

To art historians, the library of an artist is an extremely meaningful source. The list of books, authors, dates of publications, dedication pages, annotations, classification adopted and other traces of reading... provide clues to discover a little further the intimacy of an artist. Which authors did Monet read? Which authors sent him their books with a dedication page? What types of literature did he prefer? How did he treat his books? Did he take care of them or, on the contrary, did he mistreat them? Did he annotate them? Did he dog-ear the pages? By answering these questions and a close examination of the books, it is possible to map Monet's mental universe, to outline his cultural horizon, to draw a part of his social network and to discover a part of its *musée imaginaire*. Even though a work of art is never a literal transcription of a text, even though the links between what an artist has seen and the work he creates are far from obvious, his books and the illustrations they contain undoubtedly constitute an inspiration. By studying his library, the aim is to learn more about the man and the artist in order to better understand his creative process and his works. As [Levaillant et all, 2010] explain: « Une source livresque permet d'identifier, de dater et d'expliciter un thème ou un motif, selon la méthode iconographique courante qui croise les indications textuelles et les comparaisons visuelles. Elle permet aussi de démontrer une « filiation », intellectuelle ou technique, et d'inscrire enfin telle ou telle œuvre dans un environnement socio-culturel en suivant une méthode iconologique frappée au sceau de l'érudition panofskyenne sans compter que les livres sont souvent des réserves d'images en tout genre ; parmi eux, notamment, les albums lus ou regardés dans l'enfance et l'adolescence, dont deux ou trois titres suffisent à

                                                                                          http://jdmdh.episciences.org

restituer un imaginaire en acte dans le processus artistique[5] ». According to this definition, libraries are new evidences that art historians can add to the genetic file of an artwork.

## III CHALLENGES OF DIGITAL LIBRARY PROCESSING

Whether genetic, iconographic or textual, whether monographic or global and contextual, there are many approaches to studying artists' libraries. Digital humanities are another way to enlighten the richness of this resource. The *Artist Libraries Project* is also characterized by interdisciplinarity, as it compels art historians, literary historians and curators working together to make accessible to researchers and the general public not only the lists of the publications in artists' libraries from the 19[th] and 20[th] centuries but also their digital version. The goal of the *Artist Libraries Project* is to build a database that contains the bibliographical information of each book – i.e. editing, illustrations, annotations, dedication pages, placement on the shelf.... The search functionality will allow to search one artist's library or to compare several. Today limited to Monet's library, this project aims to ultimately bring together many of these libraries in order to establish international and intergenerational comparisons between the *musée imaginaire*, understood in its broadest sense, of very different artists. Furthermore, by gathering the sources in one digital tool, the project the *Artist Libraries Project* opens new perspectives on the circulation of ideas, concepts, poetics and images in the art world, on their involvement in the elaboration process and the constitution of networks and groups. For example, the significance of a book for a specific artistic group or its transmission from one generation to another is measurable. As they are a channel to spread ideas and concepts, artists' book collections sketch the cultural landscape of an era and a place. In the end, a history of ideas emerges. They also make it possible to verify certain hypotheses, such as the diffusion of scientific theories on color and light in the 19[th] century. Did the Impressionists read Goethe, Chevreul or Helmholtz? Or did they have access to these theories through second-hand literature or journals? Or, on the contrary, are these publications completely absent from their shelves? Monet only owned Félix Bracquemond's treatise[6] but he subscribed to *La Revue Blanche*. Illustrations in these books are also very important because they reveal the *musée imaginaire*, in the proper sense this time. Often considered as art books due to their rich illustrations, exhibition or sales catalogs, which are very often kept in artists' libraries, are another example of the richness of libraries as a source[7]. Their mere presence, and sometimes the annotations of price or handwritten notes they contain, reveals the sensibility of an artist. For example, in the sales catalog of Degas's collection in 1918[8], Monet had written the auction price of a painting by El Greco and Delacroix, thus leaving a hint of his artistic tastes.

Publish artists' libraries online satisfies above all a need for accessibility. Lists of books are thus no longer limited to the print edition but are accessible to everyone and everywhere. Database techniques facilitate search queries and comparison. According to *Le dictionnaire de l'École nationale supérieure des sciences de l'information et des bibliothèques*, « les bibliothèques numériques proposent de véritables collections numériques, selon une politique documentaire déterminée. Elles sont alimentées soit par des opérations de numérisation (documents patrimoniaux ou non), soit par des documents nativement numériques. Les contenus sont organisés pour en faciliter la consultation[9] ». While it partly reflects the objectives of a traditional library, this definition also suggests the many and varied possibilities offered by this new tool, including extensive consultation functionalities. A digital library is therefore not just a copy of a physical library; it is a continuation of it and even more a real extension. This statement is in line with the reflections espoused for several years by the philosopher Pierre Lévy, for whom the virtual is what exists in power and not in action. It means that the virtual world is not opposed to reality but to the present: « La virtualisation n'est pas une déréalisation (la transformation d'une réalité en un ensemble de possibles), mais une mutation d'identité, un déplacement du centre de gravité ontologique de l'objet considéré : au lieu de se définir principalement par son actualité (une « solution »), l'entité trouve désormais sa consistance essentielle dans un champ problématique. Virtualiser une entité quelconque consiste à découvrir une question générale à laquelle elle se rapporte, à faire muter l'entité en direction de cette interrogation et à redéfinir l'actualité de départ comme une réponse à une question particulière[10]. » Despite his enthusiasm for digital technology, the sociologist Michel Wieviorka has reservations about virtualization of thought and warns against the risk of « appauvrissement théorique face à une approche exagérément quantitative de questions historiques, sociologiques, anthropologiques[11] ». In addition to this theoretical uncertainty, the use of digital technology poses the challenge of practical implementation. Researchers in the humanities and social sciences have to deal with technologies that they often do not master and with project logic that is quite different from traditional research practices. We will examine a number of these questions through the example of implementing the *Artist Libraries Project*.

---

[5] « A book source makes it possible to identify, date and explain a theme or a motif, according to the current iconographic method that crosses textual indications and visual comparisons. It also makes it possible to demonstrate an intellectual or technical "filiation" and to inscribe a particular work in a socio-cultural environment by following an iconological method stamped with the seal of Panofskyan scholarship, not to mention that books are often reserves of images of all kinds; among them, in particular, albums read or viewed in childhood and adolescence, of which two or three titles are sufficient to restore an imaginary in act in the artistic process », Françoise LEVAILLANT, Jean-Roch BOUILLER et Dario GAMBONI (dirs.), *Les bibliothèques d'artistes : XXe-XXIe siècles*, Paris, PUPS, 2010, p. 12.
[6] Félix Bracquemond, *Du dessin et de la couleur*, Paris, G. Charpentier, 1885.
[7] The research project Artl@s studies exhibition catalogues by putting online a global database linked to mapping and analysis tools.
[8] *Catalogue des tableaux modernes et anciens - aquarelles, pastels, dessins par [...] composant la collection de Edgar Degas*, Paris, galerie Georges Petit, vente des 26-27 mars 1918.

## IV IDENTIFY THE TOOLS

One of the contributions of the digital humanities is the strengthening of the link between the tool and its use. Indeed, the search for suitable digital tools requires in-depth reflection on the objective of a given study. Moreover, building the database architecture involves questioning the nature of its corpus, its characteristics and its limits. An effort to define these elements must be the first step of such a project. What kind of documents are we processing? Transcribed texts, digitized texts, photographies, maps, plans, video material, audio material... Where do they come from? Who will consult them? Is the database likely to evolve or is it static? Should the database be interoperable? Is there an indexing vocabulary to describe our data? Is it a homogeneous or heterogeneous corpus? etc. In our case, the corpus is simultaneously open and close. Open because we don't know yet all the libraries we will integrate into the database but close because it will always be bibliographical data: on the one hand, bibliographic records of the books of each artist's library, which mentions the transcription of handwritten dedications and all characteristics of a copy and, on the other hand, the bibliographical description of the digitized version of the books. Built on Monet's case, the digital tool as part of the *Artist Libraries Project* has to be reliable enough to support a large number of libraries but also flexible enough to adapt to the specificities of each. Our database is built on qualified Dublin Core vocabulary[12], a descriptive format that initially included fifteen fields to describe any type of document. Created in 1995, the Dublin Core has now been extended to fifty-five descriptive fields. The use of this standard makes the system interoperable, i.e. the bibliographic records we create can be retrieved by other systems and conversely, records created by other libraries can be retrieved. We also use RAMEAU indexation language[13] to describe the books. To give access to digital books, we choose not to scan or host the books stored in the libraries because they are often common books which are already digitalized by numerous projects such as www.gallica.bnf.fr or www.openlibrary.org. Moreover, systematic digitalization as part of an academic project is often too expensive. We recovered the digital books mostly via exportable readers. On the contrary, we will photograph the marks that individualize the copy (handwritten dedication, notes, erasures, bookmarks…) and make them accessible on the website.

---

[9] "Digital libraries offer real digital collections, according to a defined documentary policy. They are supplied either by digitization operations (heritage documents or not) or by natively digital documents. The contents are organized to facilitate consultation", http://www.enssib.fr/le-dictionnaire/bibliotheques-numeriques (consulted on the 01/21/2016).

[10] "Virtualization is not a derealization (the transformation of a reality into a set of possibilities), but a change of identity, a shift in the ontological centre of gravity of the object under consideration: instead of being defined mainly by its actuality (a "solution"), the entity now finds its essential consistency in a problem field. Virtualizing any entity consists of discovering a general question to which it relates, mutating the entity towards that question and redefining the initial topicality as an answer to a particular question", Pierre Lévy, *Qu'est-ce que le virtuel ?*, Paris, 1995, p. 15.

[11] « theoretical impoverishment in front of an overly quantitative approach to historical, sociological, anthropological questions » Philippe Testard-Vaillant, « Interview de Michel Wieviorka : les sciences humaines et sociales à l'ère numérique », in *CNRS Le journal*, published online on January 10, 2014: https://lejournal.cnrs.fr/articles/interview-de-michel-wieviorka-les-sciences-humaines-et-sociales-a-lere-numerique (consulté le 21 janvier 2019).

Figure 2. Dublin Core form to add a new book on the database

---

The origin and the type of collections may also raise copyright issues and influence the level of security that must be applied to them. For example, heritage documents need a specific protection from being downloaded. In our project, the bibliographical data are not sensitive but the photographs of handwritten marks can be if the copyright is still in force. That's why they will be protected from being downloaded. The type of the original document may also affect the important issue of storage. Files of transcribed text are not very big and therefore do not require very much storage space. Conversely, images, especially in high quality - which is always desirable so that the user can "zoom in" - take up a great deal of space. It is consequently necessary to find a web host that has storage and flow management capacities adapted to the project. The web host must also provide an appropriate dissemination platform. In France, the number of academic projects with a digital component has grown quickly in recent years, but no designated institution for hosting them existed until 2013. Since then, Very Large Research Infrastructure ("Très Grande Infrastructure de Recherche", TGIR) Huma-Num aims to meet this expectation. According to the Huma-Num website, "Huma-Num aims at supporting research communities by providing services, assessment and tools on digital research data. Huma-Num coordinates the production of digital data while offering a variety of platforms and tools for the processing, conservation, dissemination and long-term preservation of digital research data[14]". Organized in consortia, it is intended to ensure not only the technical follow-up of projects but also to offer spaces for exchange and mutual assistance around common themes and objects. The *Artist Libraries Project* has been using TGIR services since the latter signed an agreement in 2016 with the labex *Les Passés dans le present*.

Once these prerequisites were established and the criteria defined, we then needed to choose an appropriate data management system. The real interdependence between IT tools and the object under examination appears once more during this phase. We required a system that could not only store and manage bibliographical data but also compare them and distinguish content specific to a book, such as the illustrations it contains. This corresponds to a flat format, unlike archive management, which requires a hierarchical format, since it deals with boxes, containing files that themselves contain documents. Numerous software programs are available to manage digital libraries. Many require payment while some are free. After a long search for the most suitable management tool for the *Artist Libraries Project*, the choice was made to use the free software Omeka, developed by the Roy Rosenzweig Center for History and New Media at George Mason University (Fairfax, VA), which meets many of our criteria[15]. Omeka is a Content Management System (CMS), a web publishing platform which manages digital content – in our case bibliographic records, linked to the digitized version of books. These data are organized by collections – each a different library – and we can consult them by list or via a search engine. While it performs the two functions, data management and dissemination remain well separated. The design or the different features can evolve or even change completely, and the content is not compromised. Omeka also has the advantage of being built on Dublin Core vocabulary. In addition to being free, open-source software benefits from a large community of users who run blogs, post tutorials online and respond via forums to technical problems that may arise.

---

[14] https://www.huma-num.fr/about-us. Consulted on January 31, 2018.
[15] https://omeka.org/

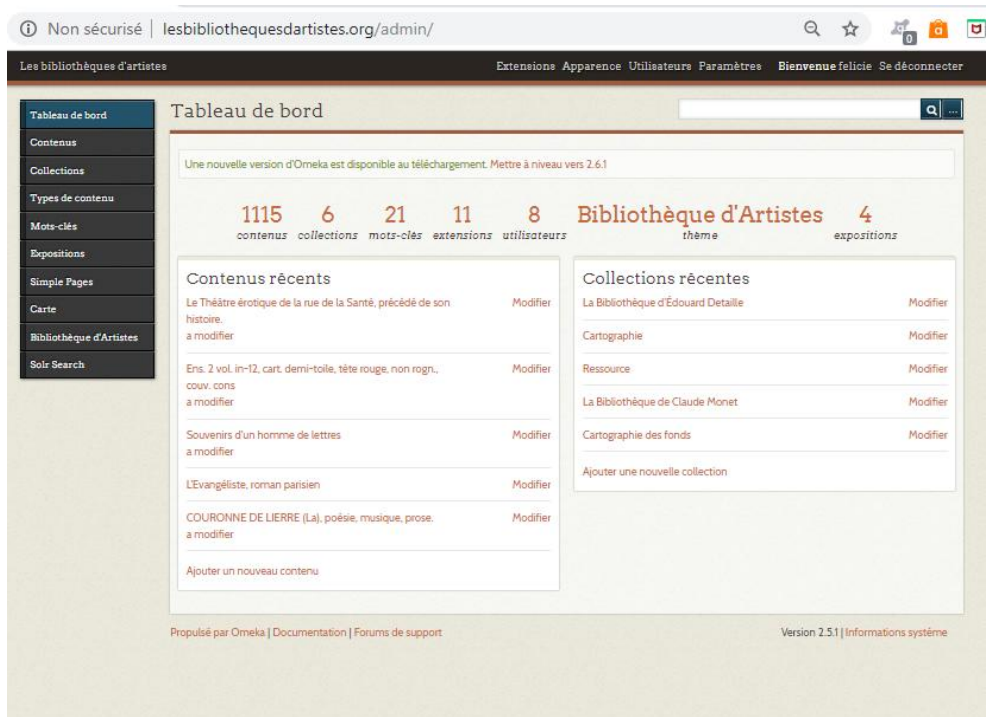

Figure 3. Dashboard of the omaka back office of www.lesbibliothequesdartistes.org

Apart from the question of content management, there is also the issue of data dissemination, since the central challenge is to make this valuable resource accessible. It is therefore necessary to design a user-friendly search function and consultation interface. For this purpose we must first ask ourselves, who will be our public? Do we target a very specialized research audience or a wider audience? The underlying question, which is particularly relevant in today's research world, is that of valorization. The tools and discourses must be adapted to the intended audiences. If the project is first designed for researchers, it can also interest a broader audience. That's why we chose to design two levels of search engines: a simple one using only keywords and an advanced one which, by refining the criteria, makes it possible to find very specific documents. In the case of artist libraries, cross-questioning will be particularly interesting, because we want to know not only if Monet had a copy of *Don Quixote* or La Fontaine's *Fables*[16], but also in which edition. Plus we want to leaf through the digital copy in question, and furthermore, we want to know if other artists of his generation also owned those books and if they were the same editions illustrated by Gustave Doré.

As the project is designed for research and not just sharing resources, we needed a website that could serve as a veritable research portal, i.e. where the community of researchers on artists' libraries can get to know each other and disseminate their studies either in academic format (scientific bibliography and papers) or a format for the broader public (virtual exhibitions). In addition to this and before integrating the entire catalog of a new library, we carry out the very important task of pointing geographically these resources. This takes the form of a map, initially elaborated as a simple plug-in and now fully developed on the online version of ARCGIS. On the website of the project, we integrated a viewer of this map which is clickable to access to ARCGIS site. Each artist library is figured on the map

---

[16] He actually did have those books: Miguel de Cervantes, *L'ingénieux Hidalgo Don Quichotte de la Manche. Trad. Louis Viardot et 370 compositions de Gustave Doré*, Paris, 1869 ; Jean de La Fontaine, *Fables. Avec les dessins de Gustave Doré*, Paris, 1868.

where it is kept. That's why there are sometimes several libraries on one point, for example at Bibliothèque Kandinsky – Centre Georges Pompidou in Paris. The libraries are classified by types (material founds, reconstituted, from inventory or sales catalogs). Thanks to the various tools (the plug-in) it offers, Omeka is a very adaptable CMS that can support these different goals. Alongside the database and its connected visualization tools, the project functions as an entire website.

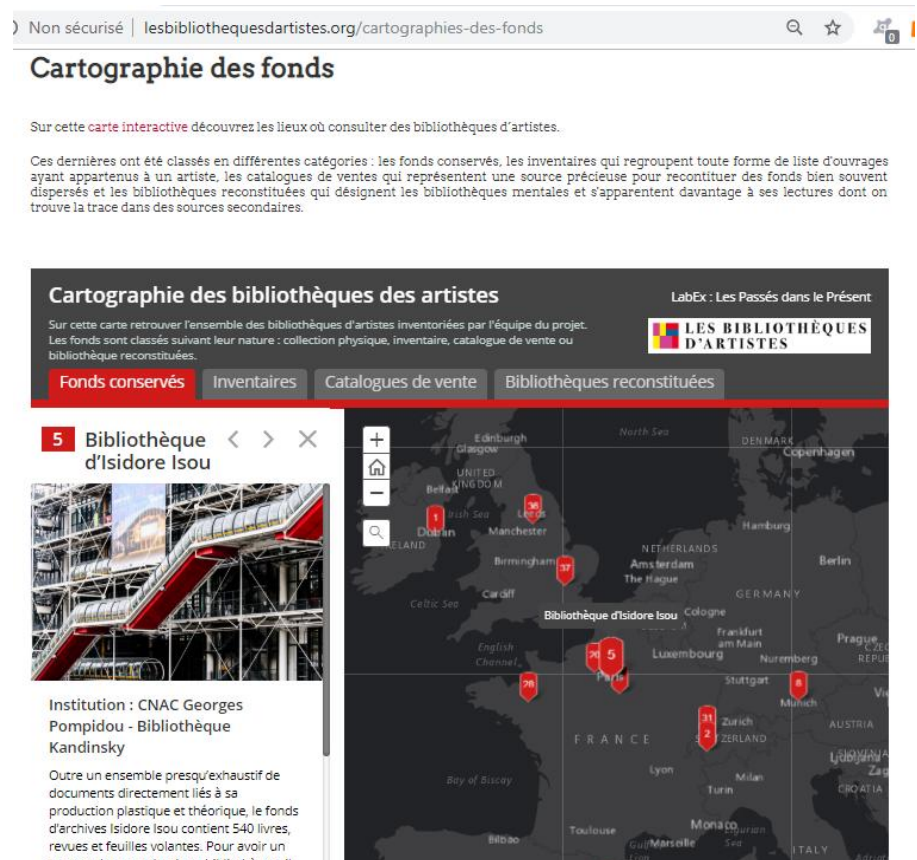

Figure 4. Map of artist libraries on www.lesbibliothequesdartistes.org

## V THE PROJECT NOW

The work on Monet's library is now finished. The catalog of the Giverny library was produced by Claudette Lindsey and Monet's foundation team upstream . It was then finalized on the occasion of the publication of the anthology of texts [Le Men et all, 2013]. A second step has consisted of checking this list of works against those kept at the Bibliothèque nationale de France and already digitized on Gallica. If they are not, the aim is to include these books in its digitization process. This phase has required more time than expected. This is partly due to the nature of the works kept in Monet's library. A large part of these books are best-selling publications of the time, which were subsequently reprinted or reissues many times, sometimes in close succession. Because we want to know what the artist actually read and saw, we normally consider only the right edition. But it is also important that the user can at least read the text. That's why, when the right digital version doesn't exist, we give access to an approximated edition. The most important issue is to clarify each case: is the digital version exactly the same as the one the artist had, or an approximate version? The time needed to achieve this work is also due to technical reasons. Because of the accuracy we need in our bibliographical matching between our primary resource and the digital version, it was not possible to know automatically if Monet's books were already digitized by Gallica. We had to check all 900 books manually[17]!

After cleaning data and choosing digital tools, the project then entered the implementation phase. All of Monet's books are now available online. In addition to its obvious scientific intrigue, this first case study was used to test and validate the choices we made, to be able to apply them to other libraries and other artists, whether painters, sculptors, architects, etc. The choice of Monet offers many advantages: first the amounts of books. 900 are enough to test the digital tools on a large scale but not too much when you have to correct some bugs. Our knowledge of the subject was an asset too. Because if we built a general database in which every library can be catalog, each one is in fact specific. For that reason, Monet's library was not a good choice for the question of reading marks. Monet almost never annotated his books. If we have still taken in account this crucial aspect of a private library, we have never managed it in the full magnitude it can have.

## VI THE PROJECT IN THE FUTURE

The shift from a monographic scale to a global scale will create the right conditions for developing analytical and critical thinking about artists' libraries, both considered as an entity in its own right and as a set of independent objects, themselves composed of pages sometimes annotated and illustrated. By comparing the libraries, the goal is no longer only to reveal the reader's portrait behind that of the artist, his interactions with the cultural sphere of his time and past, and a part of his creative process, but also the cultural environment of a time. In 2019, the libraries of the painter Edouard Detaille (1848-1912), Hans Hartung (1904-1989) and his also painter wife Anna-Eva Bergman (1909-1987), the sculptors Alberto Giacometti (1901-1966) and perhaps Antoine Bourdelle (1861-1929) will be included in the database and disclosed on the website. We also continue to fuel the map.

The developments envisaged are not only about resources but also technique. One of our major works in progress is on the illustrations contained in the books. Originally, one of the project goals was to discover a lesser known part of the artists' *musée imaginaire* built on those illustrations. We wanted to identify all of these illustrations. There are far too many of them to do manually. We thought that it would be quite easy by working on the digital versions. But when Gallica scans the books, they don't differentiate between text and picture. That's why it is impossible at this time to process automatically. The question of automatic recognition is a topical interest at the BnF so that we can hope that this will be developed soon. It may also be interesting to set up customization, i.e. the user can organize and structure the data in his own way. This would allow them to create their own thematic library, take notes or directly annotate the works, and then share the results.

Although we will surely encounter other technical challenges as we progress, we are confident that the goals of data expansion to enable a comparative approach will leaving lasting impacts on the way art history research is conducted.

---

[17] This work was conducted, with the author's oversight, by two students from Nanterre University, Hortense Rolland and Rivka Susini.